\documentclass[12pt]{iopart}

\usepackage{amssymb}
\usepackage{txfonts}
\usepackage{graphicx}
\begin{document}

\title[Materials design of Kitaev spin liquids beyond the Jackeli-Khaliullin mechanism]{Materials design of Kitaev spin liquids beyond the Jackeli-Khaliullin mechanism}

\author{Yukitoshi Motome, Ryoya Sano, Seonghoon Jang, Yusuke Sugita, and Yasuyuki Kato}

\address{Department of Applied Physics, University of Tokyo, Bunkyo, Tokyo 113-8656, Japan}
\ead{motome@ap.t.u-tokyo.ac.jp}
\vspace{10pt}

\begin{abstract}
The Kitaev spin liquid provides a rare example of well-established quantum spin liquids in more than one dimension.
It is obtained as the exact ground state of the Kitaev spin model with bond-dependent anisotropic interactions. 
The peculiar interactions can be yielded by the synergy of spin-orbit coupling and electron correlations for specific electron configuration and lattice geometry, which is known as the Jackeli-Khaliullin mechanism. 
Based on this mechanism, there has been a fierce race for the materialization of the Kitaev spin liquid over the last decade, but the candidates have been still limited mostly to $4d$- and $5d$-electron compounds including cations with the low-spin $d^5$ electron configuration, such as Ir$^{4+}$ and Ru$^{3+}$. 
Here we discuss recent efforts to extend the material perspective beyond the Jackeli-Khaliullin mechanism, by carefully reexamining the two requisites, formation of the $j_{\rm eff}=1/2$ doublet and quantum interference between the exchange processes, for not only $d$- but also $f$-electron systems. 
We present three examples: the systems including Co$^{2+}$ and Ni$^{3+}$ with the high-spin $d^7$ electron configuration, Pr$^{4+}$ with the $f^1$-electron configuration, and polar asymmetry in the lattice structure. 
In particular, the latter two are intriguing since they may realize the antiferromagnetic Kitaev interactions, in contrast to the ferromagnetic ones in the existing candidates. 
This partial overview would stimulate further material exploration of the Kitaev spin liquids and its topological properties due to fractional excitations. 
\end{abstract}

%
%
%
%
%

\section{Introduction}
\label{sec:intro}

Frustration is a key concept to open a path to the quantum spin liquid (QSL)~\cite{Lacroix2011,Diep2013}. 
The QSL is a quantum disordered state that is realized when any conventional magnetic ordering is suppressed by strong frustration in competing interactions between the magnetic moments. 
Since the proposal by P. W. Anderson in 1973~\cite{Anderson1973}, there have been extensive studies from both theoretical and experimental points of view, mostly for antiferromagnets on geometrically-frustrated lattice structures, e.g., triangular, kagome, and pyrochlore~\cite{Balents2010,Zhou2017}. 
In these systems, the exchange energy of the antiferromagnetic (AFM) Heisenberg interactions cannot be optimized on the local triangular unit, and such a frustration effect may extend to the entire lattice and suppress the formation of long-range ordering. 
Although several important aspects of the QSL, such as topological order and fractional excitations, have been unveiled thus far~\cite{Read1989,Wen1991,Senthil2000,Oshikawa2006}, it remains elusive to fully understand the physics behind, mainly due to the limited number of candidate materials and the lack of well-established theoretical tools. 

\begin{figure}[t]
\begin{center}
\includegraphics[width=0.8\columnwidth,clip, bb = 0 0 549 279]{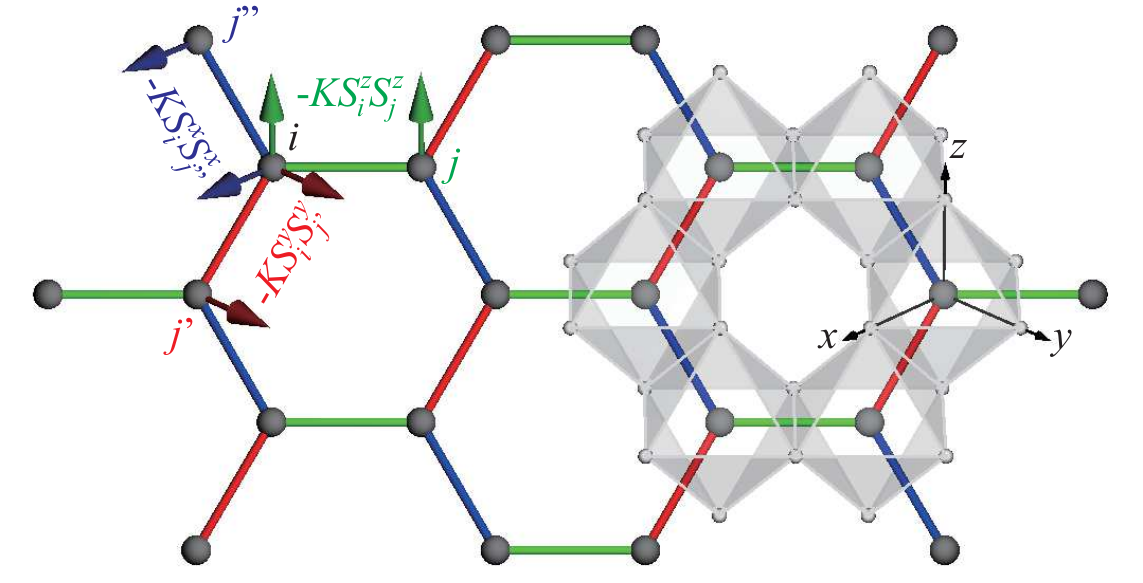}
\caption{
Schematic picture of the Kitaev model in (\ref{eq:H_Kitaev}) and its realization in an edge-sharing network of ligand octahedra. 
The colored arrows represent spins and their competition under the bond-dependent Kitaev interactions.
The black arrows are the Cartesian axes used for the definitions of $d$ and $f$ orbitals.
}
\label{fig:honeycomb}
\end{center}
\end{figure}

The Kitaev spin model, which was proposed by A. Kitaev in 2006~\cite{Kitaev2006}, brought about breakthrough in this situation. 
The model includes only nearest-neighbor interactions between spin-1/2 moments on a honeycomb structure, whose Hamiltonian is given by 
\begin{equation}
{\cal H} = - K \sum_\gamma \sum_{\langle i,j \rangle_\gamma} S_i^\gamma S_j^\gamma, 
\label{eq:H_Kitaev}
\end{equation}
where $S_i^\gamma$ denotes the $\gamma$ component of the spin-1/2 operator at site $i$, the sum of $\langle i,j \rangle_\gamma$ is taken for one of three types of nearest-neighbor bonds on the honeycomb structure ($\gamma=x,y,z$), and $K$ is the coupling constant (see figure~\ref{fig:honeycomb}). 
Since the honeycomb structure is bipartite, the model in (\ref{eq:H_Kitaev}) is free from geometrical frustration. 
However, it suffers from another type of frustration due to the bond-dependent anisotropic interactions; the spin components of the Ising-type anisotropic interactions are all different for three types of bonds on the honeycomb structure, whose energy cannot be optimized simultaneously as schematically shown in figure~\ref{fig:honeycomb}. 
Indeed, the classical counterpart of this model has macroscopic degeneracy in the ground state~\cite{Baskaran2008}. 
In the quantum case, however, the ground state of the model is exactly obtained as a QSL and that the elementary excitations are described by fractional quasiparticles, itinerant Majorana fermions and localized $Z_2$ fluxes~\cite{Kitaev2006,Baskaran2007}. 
As the model and the exact solution can be straightforwardly extended to any tricoordinate structures in any spatial dimensions, they provide rare examples of the well-established QSLs in more than one dimension~\cite{Yang2007,Yao2007,Mandal2009,Hermanns2014,O'Brien2016}. 

Bond-dependent anisotropic interactions often appear in the systems with strong entanglement between spin and orbital degrees of freedom. 
Classic examples can be found in the Kugel-Khomskii mechanism, where strong electron correlations in multiorbital systems lead to bond-dependent spin interactions through the orbital ordering and fluctuations~\cite{Kugel1973,Kugel1975}. 
Similar but different mechanism was studied for correlated electron systems with substantial strength of the relativistic spin-orbit coupling~\cite{Khaliullin2005}. 
This idea was developed for materializing the Kitaev spin model by G. Jackeli and G. Khaliullin in 2009~\cite{Jackeli2009}. 
They argued two requisites for realizing the Kitaev-type interactions: 
(i) the formation of doublet with the effective angular momentum $j_{\rm eff}=1/2$ under the crystal field and the spin-orbit coupling, and 
(ii) suppression of the conventional Heisenberg interactions due to quantum interference between different exchange processes via ligand ions. 
Under these conditions, the leading contribution to the exchange interactions between the $j_{\rm eff}=1/2$ moments is dominantly given by ferromagnetic (FM) bond-dependent interactions of Kitaev type. 
Jackeli and Khaliullin pointed out that these two requisites are potentially satisfied in $4d$- and $5d$-electron compounds with the low-spin $d^5$ electron configuration and the edge-sharing network of the ligand octahedra (see figure~\ref{fig:honeycomb}). 

Stimulated by this Jackeli-Khaliullin mechanism, there has begun a fierce race for exploration of the candidate materials for the Kitaev QSL --- see reviews, e.g., in \cite{Trebst2017preprint,Winter2017,Hermanns2018,Knolle2019,Takagi2019,Motome2020}. 
A prime candidate is honeycomb iridium oxides $A_2$IrO$_3$ with $A$=Na, Li, and Cu~\cite{Chaloupka2010,Singh2010,Singh2012,Comin2012,Foyevtsova2013,Sohn2013,Katukuri2014,Yamaji2014,Chun2015,Winter2016}. 
There are also related compounds $A'_3$LiIr$_2$O$_3$ with $A'$=Ag~\cite{Todorova2011}, Cu~\cite{Roudebush2016}, and H~\cite{Kitagawa2018}. 
Another strong candidate is a ruthenium trichloride $\alpha$-RuCl$_3$~\cite{Plumb2014,Kubota2015,Yadav2016,Winter2016,Sinn2016}, which has recently been attracting great interest owing to the discovery of half-quantized thermal Hall conductivity in a magnetic field suggesting a topological state in terms of the Majorana fermion excitations~\cite{Kasahara2018,Yokoi2020preprint}. 
There were also found three-dimensional candidates, $\beta$- and $\gamma$-Li$_2$IrO$_3$ with the hyperhoneycomb and stripy-honeycomb structures, respectively~\cite{Modic2014,Takayama2015}. 

In these candidates, the magnetic cations Ir$^{4+}$ and Ru$^{3+}$ have low-spin $d^5$ electron configurations, which can carry the $j_{\rm eff}=1/2$ Kramers doublet in the requisite (i) in the Jackeli-Khaliullin mechanism. 
Also, they share locally-tricoordinated lattice structures composed of edge-sharing ligand octahedra approximately satisfy the requisite (ii) (see figure~\ref{fig:honeycomb}). 
Nevertheless, the Kitaev QSL has not been identified in their lowest-temperature states; most of the materials exhibit magnetic long-range orders at low temperature. 
This is attributed to other non-Kitaev interactions, such as the Heisenberg exchange interaction, due to deviations from the ideal situations, such as trigonal distortions of the octahedra, and other perturbation processes~\cite{Chaloupka2010,Chaloupka2013,Rau2014}. 
In addition, in all the candidates thus far, the Kitaev interactions are thought to be FM ($K>0$) as expected from the Jackeli-Khaliullin mechanism. 
Recently, however, there were a lot of attention to the AFM Kitaev model with $K<0$, as it appears to exhibit another QSL phase in a magnetic field~\cite{Zhu2018,Gohlke2018,Nasu2018,Liang2018,Hickey2019,Ronquillo2019}. 
For further exploration of the candidates for the Kitaev QSLs including the AFM case, it is worth exploring another mechanism for Kitaev-type interactions. 

In this article, we discuss some attempts to go beyond the Jackeli-Khaliullin mechanism. 
While there have been several different pathways by using, e.g., cold atoms~\cite{Duan2003,Micheli2006,Gorshkov2013}, superconducting circuits~\cite{You2010,Kells2014,Sameti2019}, metal organic frameworks~\cite{Yamada2017a,Yamada2017b}, and Majorana cooper-pair boxes~\cite{Barkeshli2015preprint,Sagi2019}, we focus on the recent studies on inorganic materials by the authors. 
We carefully examine the two requisites for not only $d$- but also $f$-electron systems, and try to find other cases that may yield the Kitaev-type anisotropic interactions. 
Specifically, we discuss three cases. 
One is the systems with high-spin $d^7$ electron configuration, which may carry the $j_{\rm eff}=1/2$ Kramers doublet similar to the low-spin $d^5$ case~\cite{Liu2018,Sano2018}. 
The second attempt is for $f^1$-electron systems, which may also satisfy similar conditions~\cite{Jang2019,Jang2019preprint}. 
The last one is to introduce polar asymmetry with respect to the perpendicular direction to the honeycomb plane that hampers the quantum interference in the requisite (ii)~\cite{Sugita2019preprint}. 
Among the three, the latter two are intriguing since they lead to dominant AFM Kitaev interactions, which are difficult to realize in the Jackeli-Khaliullin mechanism.

The structure of this article is as follows. 
In section~\ref{sec:j=1/2}, we examine the requisite (i) in the Jackeli-Khaliullin mechanism in both $d$- and $f$-electron systems. 
We show several electron configurations which can host the $j_{\rm eff} = 1/2$ Kramers doublet. 
In section~\ref{sec:exch}, we turn to the requite (ii) and examine the relevant exchange processes. 
For the $d$-electron systems, we give an overview on several exchange interactions in the low-spin $d^5$ case discussed in the Jackeli-Khaliullin mechanism, which are common to the high-spin $d^7$ case. 
Meanwhile, among many relevant electron configurations for the $f$-electron systems, we focus on the $f^1$ case and discuss the relevant exchange interactions. 
In section~\ref{sec:beyond_JK}, we present three examples which possibly host the Kitaev coupling beyond the Jackeli-Khaliullin mechanism: the high-spin $d^7$ case, the $f^1$ case, and the polar asymmetric case. 
Finally, section~\ref{sec:summary} is devoted to the summary and perspective.

\section{$j_{\rm eff}=1/2$ Kramers doublet}
\label{sec:j=1/2}

In this section, we discuss the atomic multiplet structures of $d$- and $f$-electron configurations under Coulomb interactions, the octahedral crystal field, and the spin-orbit coupling.
Focusing on the cases with odd numbers of electrons in the outermost $d$ or $f$ shell, we present which electron configurations can host the $j_{\rm eff} = 1/2$ Kramers doublet in the lowest-energy state, compatible with the requisite (i) in the Jackeli-Khaliullin mechanism.

\subsection{$d$-electron manifold}
\label{sec:j=1/2_d}

In $d$-electron systems, one can assume that the spin-orbit coupling is weaker compared to the octahedral crystal field splitting. 
The octahedral crystal field splits the atomic $d$ levels with 10-fold degeneracy into the lower-energy $t_{2g}$ levels with sixfold degeneracy (three orbitals $\times$ spin-$1/2$) and the higher-energy $e_g$ levels with fourfold degeneracy (two orbitals $\times$ spin-$1/2$). 
Coulomb interactions are in the same order or larger compared to the octahedral crystal field splitting between the $t_{2g}$ and $e_g$ manifolds, while they become weaker when moving from $3d$ to $5d$. 
On the other hand, the spin-orbit coupling becomes stronger from $3d$ to $5d$. 
As a consequence, the energy scales for the Coulomb interactions $U$, the crystal field splitting $\Delta$, and the spin-orbit coupling (LS coupling) $\lambda$ are typically given as $O(1)$~eV, $\sim 1$~eV, and $O(0.01)$~eV, respectively, for $3d$ (namely, $U \gtrsim \Delta > \lambda$), while $O(0.1)$~eV, $\sim 1$~eV, and $O(0.1)$~eV, respectively, for $5d$ (namely $\Delta \gtrsim U \sim \lambda$). 
In the following, we discuss each $d$-electron configuration on the basis of the LS coupling scheme (the Russell-Saunders scheme). 

\begin{figure}[t]
\begin{center}
\includegraphics[width=0.6\columnwidth,clip, bb = 0 0 421 474]{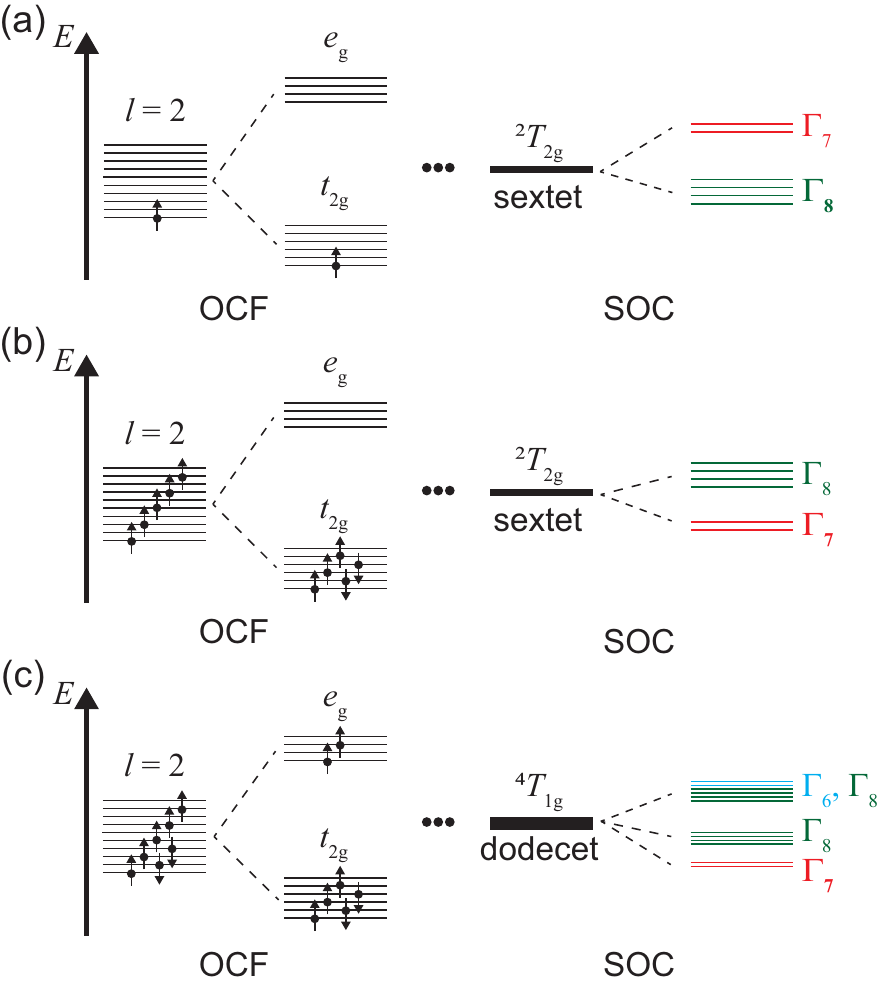}
\caption{
Schematics of the $d$-electron energy levels for (a) $d^1$, (b) low-spin $d^5$, and (c) high-spin $d^7$ cases. 
The left panels represent the level splitting by the octahedral crystal field (OCF) in the single-electron pictures, while the right panels show that by the spin-orbit coupling (SOC) in the multi-electron pictures.
}
\label{fig:d_levels}
\end{center}
\end{figure}

The simplest case is the $d^1$ case. 
In this case, the single $d$ electron occupies one of the $t_{2g}$ levels. 
The $t_{2g}$ manifold is isomorphic to the $p$-orbital manifold and described by the effective orbital angular momentum $l=1$~\cite{Abragam1970}. 
Hence, the manifold with $l=1$ and spin $s=1/2$ is sextet. 
The spin-orbit coupling splits this sextet into the $\Gamma_8$ quartet with the effective angular momentum $j_{\rm eff}=3/2$ and the $\Gamma_7$ doublet with $j_{\rm eff}=1/2$. 
In the $d^1$ case, the $\Gamma_8$ state has a lower energy than the $\Gamma_7$ state, and hence, the ground-state manifold is quartet [see figure~\ref{fig:d_levels}(a)].

In the $d^3$ case, three electrons occupy the $t_{2g}$ levels with aligning their spins and form the total spin-$3/2$ state. 
The spin-orbit coupling is ineffective as the orbital degree of freedom is quenched. 
The ground-state manifold is quartet. 

In the $d^5$ case, we may have either high-spin or low-spin state depending on the relative strength of the Coulomb interactions (Hund's coupling) to the crystal field splitting between the $t_{2g}$ and $e_g$ manifolds. 
In general, the former dominates the latter in $3d$-electron systems, while the situation might be opposite in $5d$; they may be comparable in $4d$. 
In the high-spin $d^5$ state, spins of five electrons are aligned in parallel to form the total spin $5/2$; three out of the five occupy the $t_{2g}$ levels and the rest two occupy the $e_g$ levels. 
Then, the orbital degree of freedom is quenched and the spin-orbit coupling is ineffective, as in the $d^3$ case. 
The ground-state manifold is sextet. 
On the other hand, in the low-spin $d^5$ state, all five electrons reside in the $t_{2g}$ levels. 
This situation is the electron-hole counterpart of the $d^1$ case within the $t_{2g}$ manifold. 
In this case, however, the spin-orbit coupling changes its sign, and hence, the $\Gamma_7$ doublet with  $j_{\rm eff}=1/2$ has a lower energy than the $\Gamma_8$ quartet with $j_{\rm eff}=3/2$~\cite{Jackeli2009} [see figure~\ref{fig:d_levels}(b)]. 
The $j_{\rm eff}^z=\pm 1/2$ states comprise a time-reversal Kramers pair as
\begin{equation}
\Big| j_{\rm eff}^z =\pm \frac12 \Big\rangle =
\frac{1}{\sqrt{3}} \bigg( \Big| l^z=0, s^z=\pm \frac12 \Big\rangle 
- \sqrt{2} \Big| l^z=\pm 1, s^z=\mp \frac12 \Big\rangle \bigg),
\label{eq:d5_jeff=1/2}
\end{equation}
where
\begin{equation}
| l^z = 0 \rangle = | xy \rangle, \quad 
| l^z = \pm 1 \rangle = \frac{1}{\sqrt2} \left(| zx \rangle \pm i | yz \rangle\right), 
\end{equation}
with the $t_{2g}$-orbital bases $| xy \rangle$, $| yz \rangle$, and $| zx \rangle$; $| s^z=\pm1/2 \rangle$ denote the spin-$\pm 1/2$ states. 
Here and hereafter, the $xyz$ axes are defined as shown in figure~\ref{fig:honeycomb}. 

In the $d^7$ case, we may also have both high-spin and low-spin states. 
In the high-spin case, five electrons occupy three $t_{2g}$ and two $e_g$ states with aligning their spins, and the rest two reside in the $t_{2g}$ manifold with opposite spins. 
The ground-state manifold is 12-fold degenerate (dodecet), but the spin-orbit coupling splits it into three manifolds: doublet, quartet, and sextet. 
The lowest-energy state is given by the $\Gamma_7$ doublet, which forms the Kramers doublet similarly to the low-spin $d^5$ case above~\cite{Liu2018,Sano2018} [see figure~\ref{fig:d_levels}(c)]. 
The explicit form is given by 
\begin{eqnarray}
\Big| j_{\rm eff}^z =\pm \frac12 \Big\rangle &=&
\frac{1}{\sqrt{6}} \bigg( \sqrt{3} \Big| L^z=\mp1, S^z=\pm\frac32 \Big\rangle
- \sqrt{2} \Big| L^z=0, S^z=\pm\frac12 \Big\rangle \nonumber \\
&& \quad \ \ + \Big| L^z=\pm1, S^z=\mp\frac12 \Big\rangle \bigg),
\label{eq:d7_jeff=1/2}
\end{eqnarray}
where $L$ and $S$ denote the total orbital angular momentum and the total spin, respectively. 
On the other hand, the low-spin $d^7$ case has fully-occupied $t_{2g}$ manifold and one electron in the $e_g$ manifold. 
In this case, the spin-orbit coupling is ineffective as the orbital angular momentum is quenched in the $e_g$ manifold. 
The ground state is given by quartet associated with the fourfold degeneracy of the $e_g$ manifold. 

Finally, the $d^9$ case is the electron-hole counterpart of the low-spin $d^7$ case within the $e_g$ manifold. 
Therefore, the spin-orbit coupling is ineffective and the ground-state manifold is quartet also in this case. 

\begin{table}
\caption{Ground-state manifold for different $d$-electron configurations under the octahedral crystal field. 
The degeneracy without and with the spin-orbit coupling (SOC) is shown. 
For the doublet in the low-spin $d^5$ and high-spin $d^7$ cases, the possible six-coordinate ions are also exemplified.}
\label{table_d}
\centering
\begin{tabular}{cccl}
\br
electron config. & w/o SOC & w/ SOC & possible ions \\ 
\mr
$d^1$ & sextet & quartet & \\ 
$d^3$ & quartet & & \\ 
high-spin $d^5$ & sextet & & \\ 
low-spin $d^5$ & sextet & doublet & Mn$^{2+}$, Fe$^{3+}$, Ru$^{3+}$, Os$^{3+}$, Rh$^{4+}$, Ir$^{4+}$, Pt$^{5+}$ \\
high-spin $d^7$ & dodecet & doublet & Fe$^{1+}$, Co$^{2+}$, Ni$^{3+}$ \\
low-spin $d^7$ & quartet & & \\
$d^9$ & quartet & & \\
\br 
\end{tabular}
\end{table}

The results for the $d$-electron manifold are summarized in table~\ref{table_d}. 
Among the seven cases, the Kramers doublet of our interest can appear in two cases: the low-spin $d^5$ case in (\ref{eq:d5_jeff=1/2}) and the high-spin $d^7$ case in (\ref{eq:d7_jeff=1/2}). 
The former is realized, e.g., in Ir$^{4+}$ and Ru$^{3+}$ ions as discussed in the context of the Jackeli-Khaliullin mechanism~\cite{Jackeli2009} and explored in the existing candidate materials as introduced in section~\ref{sec:intro}. 
Meanwhile, the latter is realized, e.g., in Co$^{2+}$~\cite{Liu2018,Sano2018}, which will be further discussed in section~\ref{sec:d7}.

\subsection{$f$-electron manifold}
\label{sec:j=1/2_f}

Next, let us consider $4f$-electron systems, where the LS coupling scheme is applicable. 
We will comment on $5f$-electron systems in the end of this section. 
In the $4f$ case, the spin-orbit coupling is usually larger than the octahedral crystal field; typically, $U = O(1)$~eV, $\lambda = O(0.1)$~eV, and $\Delta = O(0.01)$~eV (namely $U > \lambda > \Delta$). 
The spin-orbit coupling entangles the orbital angular momentum $l=3$ for the $f$-electron manifold and the spin angular momentum to form the multiplets. 
The degeneracy in the multiplets are lifted by the octahedral crystal field. 
The effect of the octahedral crystal field is in general described by the Hamiltonian 
\begin{equation}
{\cal H}_{\rm OCF} = B_{40} O_4 + B_{60} O_6,
\label{eq:H_OCF1}
\end{equation}
where $O_4 = O_{40} + 5O_{44}$ and $O_6 = O_{60} - 21O_{64}$ with the rank-$r$ Stevens operators $O_{rs}$ ($s=-r, -r+1, \cdots, r$); $B_{40}$ and $B_{60}$ are the coefficients. 
This is written by two parameters $W$ and $x$ as~\cite{Lea1962}
\begin{equation}
{\cal H}_{\rm OCF} = W \left\{ x \frac{O_4}{F_J(4)} + (1-|x|) \frac{O_6}{F_J(6)} \right\}, 
\label{eq:H_OCF2}
\end{equation}
where $F_J(4)$ and $F_J(6)$ are the factors estimated for each $4f$-electron configuration in \cite{Lea1962}. 
In the following, we consider the ground-state multiplet for each $4f$-electron configuration under the strong spin-orbit coupling and the octahedral crystal field in the LS coupling scheme. 

\begin{figure}[t]
\begin{center}
\includegraphics[width=1.0\columnwidth,clip, bb = 0 0 786 618]{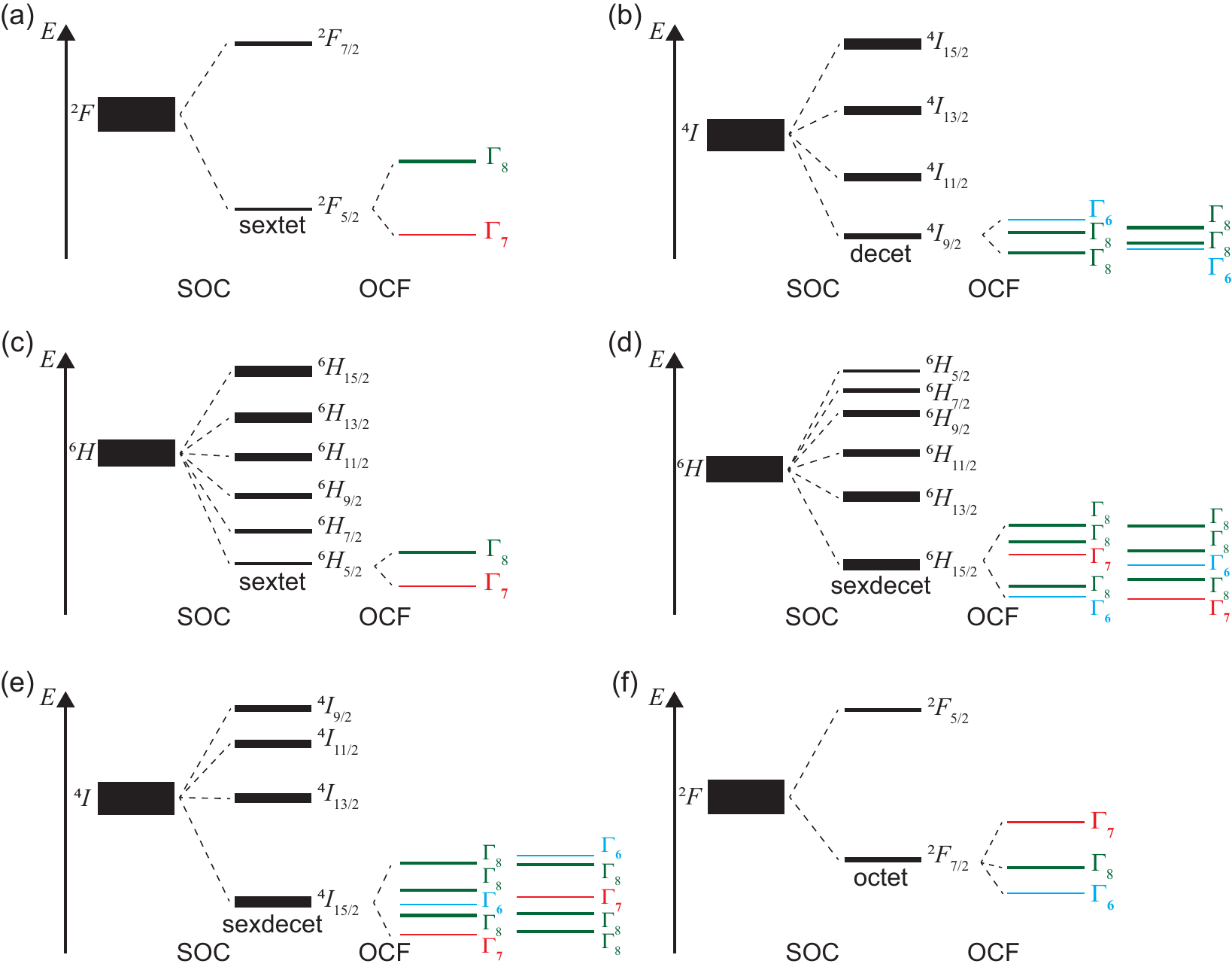}
\caption{
Schematics of the splitting of the $f$-electron energy levels by the spin-orbit coupling (SOC) and the octahedral crystal field (OCF) for (a) $4f^1$, (b) $4f^3$, (c) $4f^5$, (d) $4f^9$, (e) $4f^{11}$, and (f) $4f^{13}$ cases. 
}
\label{fig:f_levels}
\end{center}
\end{figure}

Let us start with the $4f^1$ case. 
The LS coupling scheme predicts the total angular momentum $5/2$ state with the orbital angular momentum $3$ and the spin angular momentum $1/2$, which is denoted as $^2F_{5/2}$. 
Under the octahedral crystal field in (\ref{eq:H_OCF1}), the $^2F_{5/2}$ sextet is split into the $\Gamma_7$ doublet and the $\Gamma_8$ quartet. 
This case is rather special because $x=1$ in (\ref{eq:H_OCF2}) (this is concluded from $B_{60}=0$ in (\ref{eq:H_OCF1})~\cite{Lea1962}) and $W$ is positive. 
As a consequence, the first term in (\ref{eq:H_OCF2}) always lowers the energy of the $\Gamma_7$ state compared to $\Gamma_8$ [see figure~\ref{fig:f_levels}(a)]. 
Thus, the ground-state multiplet for the $4f^1$-electron configuration is given by the $\Gamma_7$ doublet with the effective angular momentum $j_{\rm eff}=1/2$, which comprises a time-reversal Kramers pair like those in the low-spin $d^5$ and high-spin $d^7$ states~\cite{Jang2019,Jang2019preprint}. 
The explicit form of the Kramers pair is given by
\begin{equation}
\Big| j_{\rm eff}^z = \pm \frac12 \Big\rangle = 
\frac{i}{\sqrt{6}} \bigg(-\sqrt{5} \Big| j=\frac{5}{2},j^z=\mp\frac{3}{2} \Big\rangle
+ \Big| j=\frac{5}{2},j^z=\pm\frac{5}{2} \Big\rangle \bigg). 
\label{eq:4f1}
\end{equation}
This is also written in terms of the $f$-orbital bases as
\begin{equation}
\Big| j_{\rm eff}^z = \pm \frac12 \Big\rangle = 
\frac{1}{\sqrt{21}} \Big(2ic^{\dagger}_{\xi\mp} \mp 2c^{\dagger}_{\eta\mp} 
\pm 2ic^{\dagger}_{\zeta\pm} + 3c^{\dagger}_{A\pm} \Big) |0\rangle,
\end{equation}
where ($\it{\xi}$, $\it{\eta}$, $\it{\zeta}$) and $A$ denote the $f$ orbitals with the irreducible representations T$_{2u}$ and A$_{2u}$, respectively~\cite{Takegahara1980}, and $c_{\nu\sigma}^\dagger$ is a creation operator of an electron with orbital $\nu$ and spin $\sigma$ ($+$ and $-$ denote spin $\uparrow$ and $\downarrow$, respectively); $|0\rangle$ is the vacuum of $f$ electrons. 

The $4f^3$ case takes the $^4I_{9/2}$ multiplet. 
The 10-fold degeneracy is lifted by the octahedral crystal field into two $\Gamma_8$ quartets and one $\Gamma_6$ doublet. 
In this case, according to \cite{Lea1962}, $W$ is negative and $x$ is positive in (\ref{eq:H_OCF2}), which predicts that the lowest-energy multiplet is given by one of the $\Gamma_8$ quartet for $x\lesssim 0.834$ and the $\Gamma_6$ doublet for $x\gtrsim 0.834$. 
The schematics for both cases are shown in figure~\ref{fig:f_levels}(b). 
The $\Gamma_6$ doublet that we are interested in here is explicitly given by 
\begin{eqnarray}
\Big| j_{\rm eff}^z = \pm \frac12 \Big\rangle &=& 
\frac{i}{12} \bigg(\sqrt{6} \Big| j=\frac{9}{2},j^z=\mp\frac{7}{2} \Big\rangle
+ 2\sqrt{21} \Big| j=\frac{9}{2},j^z=\pm\frac{1}{2} \Big\rangle \nonumber \\
&& \quad +3\sqrt{6} \Big| j=\frac{9}{2},j^z=\pm\frac{9}{2} \Big\rangle \bigg).
\end{eqnarray}

In the $4f^5$ case, the multiplet under the spin-orbit coupling is given by the $^6H_{5/2}$ sextet. 
The sixfold manifold is lifted by the octahedral crystal field in a similar manner to the $4f^1$ case. 
We therefore end up with the $\Gamma_7$ Kramers doublet, as shown in figure~\ref{fig:f_levels}(c). 
The wave function is common to (\ref{eq:4f1}) in the $4f^1$ case. 

In the $4f^7$ case, the LS coupling scheme predicts the $^8S_{7/2}$ multiplet. 
In this case, the total angular momentum is equal to the total spin, and the orbital is quenched. 
In this situation, the crystal field is irrelevant and the system has an isotropic magnetic moment. 

Both $4f^9$ and $4f^{11}$ cases have 16-fold multiplets: $^6H_{15/2}$ for $4f^9$ and $^4I_{15/2}$ for $4f^{11}$. 
Although $W$ is positive for both cases, $x$ is negative for the former and positive for the latter~\cite{Lea1962}. 
In both cases, however, the degeneracy is lifted by the octahedral crystal field into one $\Gamma_7$ doublet, three $\Gamma_8$ quartet, and one $\Gamma_6$ doublet. 
The ground-state multiplet depends on the value of $x$; the $4f^9$ case takes $\Gamma_6$ for $x\lesssim - 0.459$ and $\Gamma_7$ for $x\gtrsim -0.459$, while the $4f^{11}$ case takes $\Gamma_7$ for $x\lesssim 0.585$ and one of $\Gamma_8$ for $x\gtrsim 0.585$~\cite{Lea1962}. 
The schematics are displayed in figures~\ref{fig:f_levels}(d) and \ref{fig:f_levels}(e). 
The $\Gamma_6$ state for the $4f^9$ case is given by 
\begin{eqnarray}
\Big| j_{\rm eff}^z = \pm \frac12 \Big\rangle &=& 
\frac{i}{24} \bigg(\pm\sqrt{195} \Big| j=\frac{15}{2},j^z=\mp\frac{15}{2} \Big\rangle
\pm3\sqrt{7} \Big| j=\frac{15}{2},j^z=\mp\frac{7}{2} \Big\rangle \nonumber \\
& & \quad \ \ \pm3\sqrt{33}\Big| j=\frac{15}{2},j^z=\pm\frac{1}{2} \Big\rangle
\pm\sqrt{21} \Big| j=\frac{15}{2},j^z=\pm\frac{9}{2} \Big\rangle \bigg),
\end{eqnarray}
while the $\Gamma_7$ one by 
\begin{eqnarray}
\Big| j_{\rm eff}^z = \pm \frac12 \Big\rangle &=& 
\frac{i}{24} \bigg(\pm\sqrt{33} \Big| j=\frac{15}{2},j^z=\mp\frac{11}{2} \Big\rangle
\pm3\sqrt{13} \Big| j=\frac{15}{2},j^z=\mp\frac{3}{2} \Big\rangle \nonumber \\
& & \quad \ \ \mp\sqrt{195} \Big| j=\frac{15}{2},j^z=\pm\frac{5}{2} \Big\rangle
\mp\sqrt{231}\Big| j=\frac{15}{2},j^z=\pm\frac{13}{2} \Big\rangle \bigg),
\end{eqnarray}
which is common to the $4f^9$ and $4f^{11}$ cases.

Finally, the $4f^{13}$ case is in the $^2F_{7/2}$ octet, which is split by the octahedral crystal field into $\Gamma_6$, $\Gamma_7$, and $\Gamma_8$. 
In this case, $W$ is positive and $x$ is negative, predicting that the ground state is given by the $\Gamma_6$ doublet, as shown in figure~\ref{fig:f_levels}(f)~\cite{Lea1962}. 
The $\Gamma_6$ state is written as
\begin{equation}
\Big| j_{\rm eff}^z = \pm \frac12 \Big\rangle = 
\frac{i}{6} \bigg(\mp\sqrt{15} \Big| j=\frac{7}{2},j^z=\mp\frac{7}{2} \Big\rangle
\mp\sqrt{21} \Big| j=\frac{7}{2},j^z=\pm\frac{1}{2} \Big\rangle \bigg). 
\end{equation}
This is also written in terms of the $f$-orbital bases as
\begin{equation}
\Big| j_{\rm eff}^z = \pm \frac12 \Big\rangle = 
\frac{1}{\sqrt{3}} \Big(-i c^{\dagger}_{\alpha\mp} \pm c^{\dagger}_{\beta\mp} \mp i c^{\dagger}_{\gamma\pm}\Big) | 0 \rangle,
\end{equation}
where ($\alpha$, $\beta$, $\gamma$) denote the $f$ orbitals with the irreducible representation T$_{1u}$ ~\cite{Takegahara1980}.

\begin{table}
\caption{Ground-state multiples for different $4f$-electron configurations. 
The multiplet with the spin-orbit coupling (SOC) and the octahedral crystal field (OCF) is shown. 
The possible six-coordinate ions are also exemplified.}
\label{table_f}
\centering
\begin{tabular}{cccl}
\br
electron config. & w/ SOC & w/ OCF & possible ions \\ 
\mr
$4f^1$ & $^2F_{5/2}$ & $\Gamma_7$ & Ce$^{3+}$, Pr$^{4+}$ \\ 
$4f^3$ & $^4I_{9/2}$ & $\Gamma_8$ or $\Gamma_6$ & Nd$^{3+}$ \\ 
$4f^5$ & $^6H_{5/2}$ & $\Gamma_7$ & Sm$^{3+}$ \\ 
$4f^7$ & $^8S_{7/2}$ & & Eu$^{2+}$, Gd$^{3+}$, Tb$^{4+}$ \\
$4f^9$ & $^6H_{15/2}$ & $\Gamma_6$ or $\Gamma_7$ & Dy$^{3+}$ \\ 
$4f^{11}$ & $^4I_{15/2}$ & $\Gamma_7$ or $\Gamma_8$ & Er$^{3+}$ \\ 
$4f^{13}$ & $^2F_{7/2}$ & $\Gamma_6$ & Tm$^{2+}$, Yb$^{3+}$ \\ 
\br 
\end{tabular}
\end{table}

The results for the $4f$-electron manifold are summarized in table~\ref{table_f}. 
All the $4f$-electron configurations, except for $4f^7$, have a chance to form the Kramers doublet. 
In the $4f^3$ and $4f^{11}$ cases, however, the system becomes doublet or quartet depending on the value of $x$. 
We note that estimates of $x$ were recently given as $x\simeq 0.743$ for the $4f^3$ case and $x\simeq 0.640$ for the $4f^{11}$ case~\cite{Duan2010}. 
On this basis, both $4f^3$ and $4f^{11}$ cases are plausibly in the $\Gamma_8$ quartet. 
Meanwhile, the value of $x$ for the $4f^9$ case was estimated as $x\simeq -0.821$, for which probably the system takes the $\Gamma_6$ doublet. 
Therefore, in the $4f$-electron cases, there are four good candidates for the Kramers doublet: $\Gamma_7$ for $4f^1$ and $4f^5$, and $\Gamma_6$ for $4f^9$ and $4f^{13}$. 
We note that the $\Gamma_7$ case with the $4f^1$-electron configuration was recently discussed for compounds with high-valent Pr$^{4+}$ ions by some of the authors~\cite{Jang2019,Jang2019preprint} (see section~\ref{sec:f1}), while the $\Gamma_6$ and $\Gamma_7$ cases with the $4f^{13}$-electron configuration were discussed for compounds with Yb$^{3+}$ ions \cite{Rau2018}. 

In the cases of $5f$ electrons, among natural elements up to Uranium, the possible Kramers doublet is limited to $5f^1$ (e.g., Pa$^{4+}$ and U$^{5+}$) and $5f^3$ (U$^{3+}$) electron configurations. 
In these cases, the Coulomb interactions can be comparable or even smaller than the spin-orbit coupling, and hence, the LS coupling scheme might be no longer valid. 
Nevertheless, in the $5f^1$ case, the ground state will be the $\Gamma_7$ multiplet with $j_{\rm eff}=1/2$, similarly to the $4f^1$ case, as the Coulomb interactions are irrelevant in this single-electron case.
On the other hand, the situation is complicated in the $5f^3$ case; 
for instance, the ground state of U$^{3+}$ was given as a mixture of $^4I$ and $^2H$, while $^4I_{9/2}$ is expected from the LS coupling scheme~\cite{Ursu1984}.
We leave this complicated situation as a future issue.

\section{Exchange interactions}
\label{sec:exch}

In this section, focusing on the electron configurations with the Kramers doublet found in section~\ref{sec:j=1/2}, we discuss the exchange processes of electrons that give effective magnetic couplings between the $j_{\rm eff}=1/2$ moments. 
In section~\ref{sec:exch_d}, we discuss the low-spin $d^5$ and high-spin $d^7$ cases (see table~\ref{table_d}). 
In section~\ref{sec:exch_f}, among several possibilities in the $f$-electron systems (see table~\ref{table_f}), we discuss the simplest case with $4f^1$ electron configuration.

\subsection{$d$ orbitals}
\label{sec:exch_d}

In the $d$-electron systems, as discussed in section~\ref{sec:j=1/2_d}, there are two possible electron configurations hosting the $j_{\rm eff}=1/2$ Kramers doublet: the low-spin $d^5$ and high-spin $d^7$ cases. 
The exchange processes in the former case with the edge-sharing network of ligand octahedra were discussed in the literatures~\cite{Khaliullin2005,Jackeli2009,Chaloupka2010,Chaloupka2013,Rau2014,Winter2016}. 
They may lead to a predominant Kitaev interaction (the Jackeli-Khaliullin mechanism). 
We briefly describe the essence of these arguments in the following. 
The latter high-spin $d^7$ case was pointed out recently to have similar exchange processes~\cite{Liu2018,Sano2018}. 
We will comment on this point in the end of this section. 

\begin{figure}[t]
\begin{center}
\includegraphics[width=1.0\columnwidth,clip, bb = 0 0 613 272]{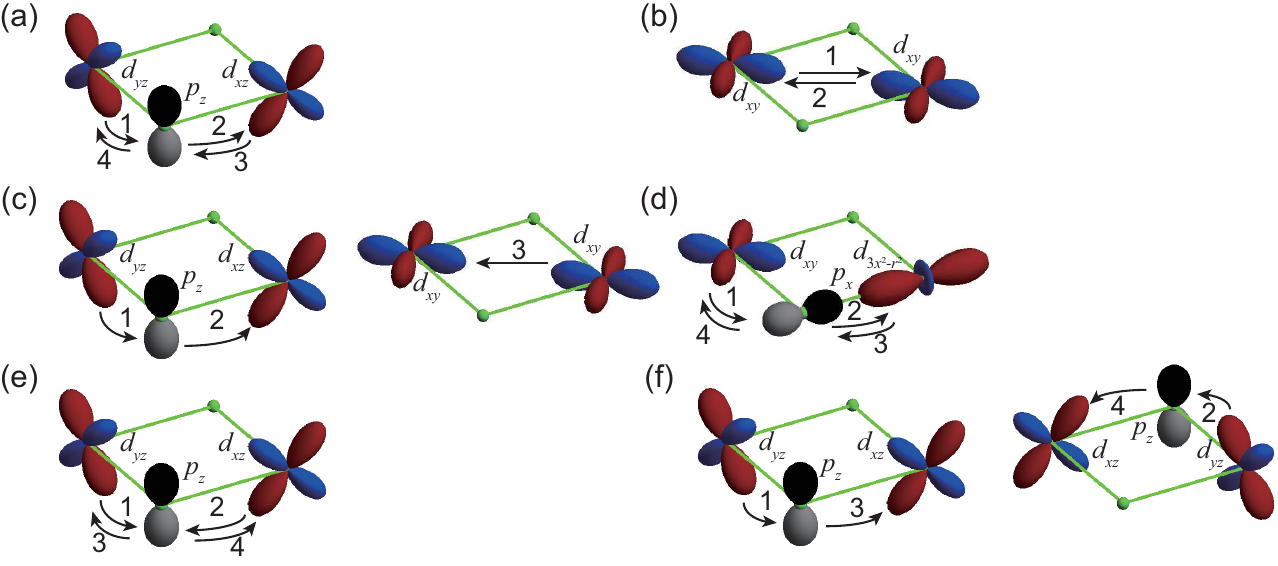}
\caption{
Schematic pictures for the exchange processes in the $d$-electron case. 
The objects with red and blue ovals represent the $d$ orbitals, and those with gray ones are the $p$ orbitals at the ligand sites. 
The numbered arrows denote the sequence of hoppings in the perturbation. 
(a) shows two indirect $d$-$p$-$d$ hoppings relevant to the FM Kitaev interaction, 
(b) two direct $d$-$d$ hoppings relevant to the AFM Heisenberg interaction, 
(c) a mixture of indirect and direct ones relevant to the $\Gamma$ interaction, 
(d) $t_{2g}$-$e_g$ hoppings, and 
(e), (f) $p$-$d$ charge transfer excitations. See the text for details. 
}
\label{fig:d_exchanges}
\end{center}
\end{figure}

In the low-spin $d^5$ case, the most relevant exchange process to the Kitaev interaction is given by the indirect hopping $t_{dpd}$ between nearest-neighbor $t_{2g}$ orbitals via ligand $p$ orbitals. 
In the edge-sharing geometry, there are two paths for the $d$-$p$-$d$ hopping, which cancel the Heisenberg interaction and leave the Kitaev interaction as the dominant contribution. 
This gives rise to an effective Hamiltonian in the form of (\ref{eq:H_Kitaev}). 
The Kitaev coupling constant $K$ is obtained by the second-order perturbation in terms of $t_{dpd}$ [see figure~\ref{fig:d_exchanges}(a)] as  
\begin{equation}
K \simeq \frac83 \frac{t_{dpd}^2}{U} \frac{J_{\rm H}}{U}, 
\label{eq:K_d5}
\end{equation}
where $U$ and $J_{\rm H}$ are the Coulomb repulsion and the Hund's coupling between $t_{2g}$ electrons at the same site, respectively. 
There are two important aspects in (\ref{eq:K_d5}): 
(i) $K$ is always positive, namely, FM, and 
(ii) it is proportional to the Hund's coupling $J_{\rm H}$. 
The latter comes from the fact that, in the intermediate state in the perturbation processes, an electron is transferred to one of the $j_{\rm eff}=3/2$ quartet at the neighboring site and feels the Hund's coupling energy for another electron in the $j_{\rm eff}=1/2$ doublet. 

In addition, there are two important exchange processes. 
One is given by the direct hopping $t_{dd}$ between neighboring $t_{2g}$ orbitals [see figure~\ref{fig:d_exchanges}(b)], which results in the Heisenberg interaction described by 
$J \mathbf{S}_i \cdot \mathbf{S}_j$ 
with 
\begin{equation}
J \simeq \frac23 \frac{t_{dd}^2}{U}. 
\label{eq:J_f}
\end{equation}
Here, $J$ is positive, namely AFM [see also (\ref{eq:H_KJG}) below]. 
The other is given by the combination of the direct and indirect hoppings [see figure~\ref{fig:d_exchanges}(c)]. 
This leads to a symmetric off-diagonal interaction, dubbed the $\Gamma$ interaction, given by 
$\Gamma (S_i^{\gamma'} S_j^{\gamma''} + S_i^{\gamma''} S_j^{\gamma'})$ on the $\gamma$ bond [$(\gamma, \gamma', \gamma'') = (x,y,z)$ and the cyclic permutations], 
with
\begin{equation}
\Gamma \simeq -\frac{16}{9} \frac{t_{dpd}t_{dd}}{U} \frac{J_{\rm H}}{U}. 
\end{equation} 
We note that there are also contributions from the $t_{2g}$-$e_g$ hopping [see figure~\ref{fig:d_exchanges}(d)] and the indirect hopping involving $p$-$d$ charge transfer excitations [see figures~\ref{fig:d_exchanges}(e) and \ref{fig:d_exchanges}(f)], but they give rise to subdominant contributions to the Kitaev and Heisenberg interactions in the same forms as shown above. 
It is worth noting that the former $t_{2g}$-$e_g$ hopping leads to an AFM Kitaev interaction, while the contribution is in general small compared to (\ref{eq:J_f}) because of the large crystal field splitting between the $t_{2g}$ and $e_g$ manifolds. 
We will return to this point in the next section~\ref{sec:exch_f}. 

Summarizing the above, the effective Hamiltonian for the $j_{\rm eff}=1/2$ moments in the low-spin $d^5$ case is well described by the Kitaev, Heisenberg, and symmetric off-diagonal interactions as~\cite{Jackeli2009, Chaloupka2010,Chaloupka2013,Rau2014} 
\begin{equation}
{\cal H} = \sum_\gamma \sum_{\langle i,j \rangle_\gamma} \left\{ 
-K S_i^\gamma S_j^\gamma 
+ J \mathbf{S}_i \cdot \mathbf{S}_j 
+ \Gamma (S_i^{\gamma'} S_j^{\gamma''} + S_i^{\gamma''} S_j^{\gamma'}) \right\}. 
\label{eq:H_KJG}
\end{equation}
More complete formulae for the coupling constants, $K$, $J$, and $\Gamma$, are found in \cite{Rau2014,Winter2016}. 
The actual values of the coupling constants vary among the candidate materials such as iridium oxides and ruthenium trichloride; for instance, refer to a review in \cite{Winter2017}. 
Note that the above arguments are for the case with ideal ligand octahedra; in real compounds, distortions of the octahedra modulate the atomic level scheme as well as the exchange processes, which lead to other contributions to the effective Hamiltonian. 
An interesting example will be discussed in section~\ref{sec:polar}. 

In the high-spin $d^7$ case, the exchange processes are similar to those in the low-spin $d^5$ case~\cite{Liu2018,Sano2018}. 
The resultant Hamiltonian is given by the same form as (\ref{eq:H_KJG}). 
An important difference from the $d^5$ case is in an additional contribution from the exchange processes between the $e_g$ electrons. 
This is absent in the $d^5$ case as the $e_g$ orbitals are empty. 
It was pointed out that this contribution leads to a FM Heisenberg interaction, which competes with the AFM one in (\ref{eq:H_KJG}) and possibly makes the system being closer to the pure Kitaev case when the $\Gamma$ term can be neglected~\cite{Liu2018}.

\subsection{$f$ orbitals}
\label{sec:exch_f}

As discussed in section~\ref{sec:j=1/2_f}, in the case of the $4f$-electron manifold, there are several different electron configurations possibly hosting  the $j_{\rm eff}=1/2$ Kramers doublet. 
Among them, the exchange processes were recently examined for the $4f^1$ case~\cite{Jang2019,Jang2019preprint} and its electron-hole counterpart, the $4f^{13}$ case~\cite{Rau2018}. 
While the latter case with the $\Gamma_6$ doublet was found to yield a dominant AFM Heisenberg interaction~\cite{Rau2018}, the former $\Gamma_7$ was shown to give an AFM Kitaev coupling, in contrast to the FM one in the $d^5$ and $d^7$ cases~\cite{Jang2019,Jang2019preprint}. 
We here briefly discuss the exchange processes in the $4f^1$ case. 

\begin{figure}[t]
\begin{center}
\includegraphics[width=0.6\columnwidth,clip, bb = 0 0 324 89]{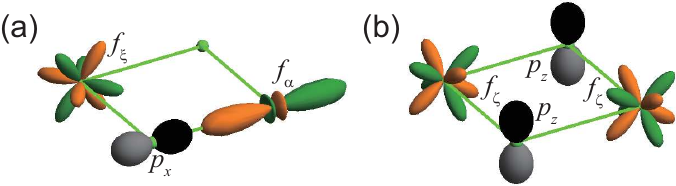}
\caption{
Schematic pictures for two types of $f$-$p$-$f$ hoppings: (a) $t_{\xi p_x \alpha}$  and (b) $t_{\zeta p_z \zeta}$. 
}
\label{fig:f_hoppings}
\end{center}
\end{figure}

In the $4f^1$ systems with edge-sharing octahedra, the dominant hoppings are two types of indirect ones $t_{fpf}$ between neighboring $f$ orbitals via ligand $p$ orbitals: $t_{\xi p_x \alpha}$ ($= -t_{\eta p_y \beta}$) and $t_{\zeta p_z \zeta}$ [see figures~\ref{fig:f_hoppings}(a) and \ref{fig:f_hoppings}(b), respectively]. 
There are also contributions from the direct ones $t_{ff}$ between the same sets of $f$ orbitals, $t_{\xi\alpha}$ ($= -t_{\eta\beta}$) and $t_{\zeta\zeta}$. 
Although there are other nonzero hoppings, let us focus on the exchange processes from these dominant ones. 

\begin{figure}[t]
\begin{center}
\includegraphics[width=1.0\columnwidth,clip, bb = 0 0 619 267]{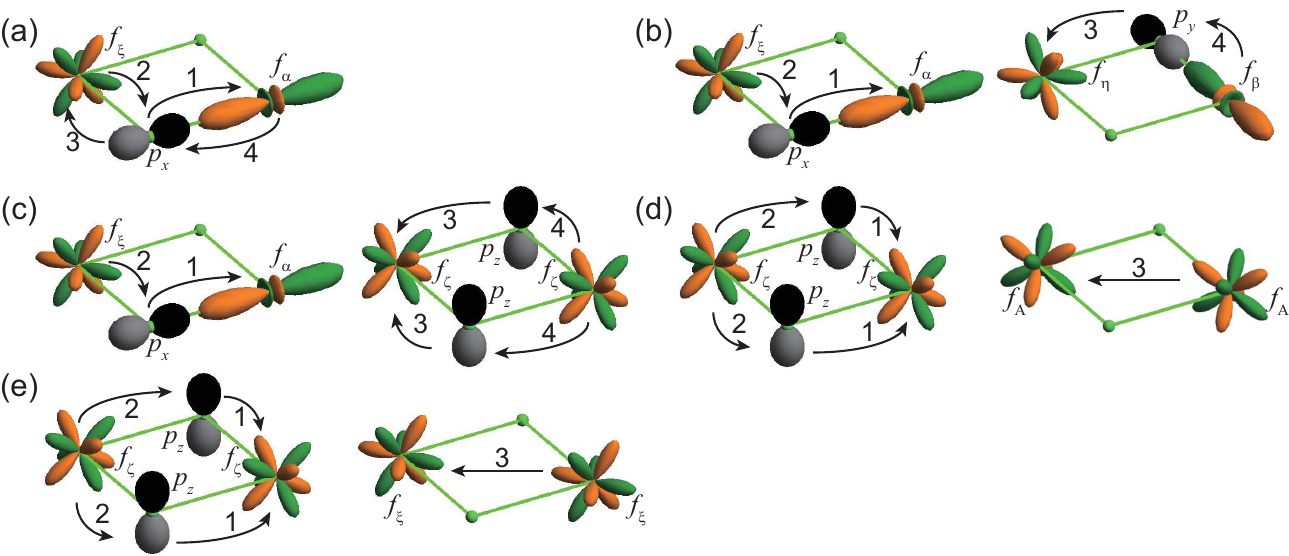}
\caption{
Schematic pictures for the relevant exchange processes in the $f$-electron case. 
The objects with orange and green ovals represent the $f$ orbitals, and those with gray ones are the $p$ orbitals at the ligand sites. 
The numbered arrows denote the sequence of hoppings in the perturbation. 
(a)-(c) show two indirect $f$-$p$-$f$ hoppings and 
(d), (e) mixtures of indirect and direct ones. See the text for details. 
}
\label{fig:f_exchanges}
\end{center}
\end{figure}

The most relevant exchange process is given by the combinations of two of the indirect and direct hoppings. 
There are five types of such exchange processes, as showcased in figure~\ref{fig:f_exchanges}: 
The former three in figures~\ref{fig:f_exchanges}(a)-\ref{fig:f_exchanges}(c) involve only the indirect ones, while the latter two in figures~\ref{fig:f_exchanges}(d) and \ref{fig:f_exchanges}(e) involve both indirect and direct ones. 
All these exchange processes contribute to the AFM Kitaev interaction; in particular, the largest contribution comes from $t_{\xi p_x \alpha}$ and $t_{\xi\alpha}$ (or $t_{\eta p_y \beta}$ and $t_{\eta\beta}$), as exemplified in figure~\ref{fig:f_exchanges}(a). 
This is in stark contrast to the $d$-electron cases in section~\ref{sec:exch_d}, where the Kitaev interaction is predominantly FM. 

The reason why the Kitaev interaction is AFM is qualitatively understood as follows~\cite{Jang2019}. 
The relevant hoppings, $t_{\xi p_x \alpha}$ and $t_{\xi\alpha}$ ($t_{\eta p_y \beta}$ and $t_{\eta\beta}$), look similar to the $t_{2g}$-$e_g$ hopping in figure~\ref{fig:d_exchanges}(d); both $f$ and $d$ cases involve the large overlap of $\sigma$ type between the $f$($d$) and $p$ orbitals. 
In the $d$-electron cases, as mentioned in section~\ref{sec:exch_d}, the $t_{2g}$-$e_g$ hopping leads to an AFM Kitaev interaction. 
Similarly, the hoppings $t_{\xi p_x \alpha}$ and $t_{\xi\alpha}$ give rise to the AFM Kitaev interaction. 
The difference between the $d$- and $f$-electron cases lies in the amplitude of the coupling constant. 
As stated in section~\ref{sec:exch_d}, in the $d^5$ and $d^7$ cases, the amplitude is much reduced by the large crystal field splitting between $t_{2g}$ and $e_g$ which enters in the energy of the intermediate state in the perturbation. 
In contrast, in the $4f^1$ case, the crystal field splitting is small, which leaves the large amplitude of the AFM Kitaev coupling. 
Thus, the AFM Kitaev interaction appears due to the spatial anisotropy of the $f$ orbitals and the small crystal field splitting. 

Besides the AFM Kitaev interaction, other exchange processes give a subdominant Heisenberg interaction. 
There are two dominant processes: One is via the indirect hopping $t_{\zeta p_z \zeta}$ and the direct one $t_{AA}$, and the other is via the two indirect hoppings $t_{\zeta p_z \zeta}$. 
The formers contribute to an AFM Heisenberg interaction, while the latter to FM. 
After the cancellation between the two, the net Heisenberg interaction becomes small and AFM. 

Therefore, the effective Hamiltonian for the $j_{\rm eff}=1/2$ moments in the $4f^1$ case is approximately given by the dominant AFM Kitaev interaction $K<0$ and the subdominant AFM Heisenberg interaction $J>0$~\cite{Jang2019,Jang2019preprint}. 
There is no contribution to the $\Gamma$ term in (\ref{eq:H_KJG}) in the ideal case. 
When considering deviations from the ideal octahedra, such as trigonal distortions, the ratio between $|K|$ and $J$ changes systematically, and in addition, other exchange interactions also come into play. 
A recent systematic study on a series of honeycomb compounds $A_2$PrO$_3$ ($A$: alkali metals) revealed that the trigonal distortions become larger for larger $A$-site ionic radii, and accordingly, $|K|$ becomes smaller while $J$ does not change significantly~\cite{Jang2019preprint}. 
At the same time, a symmetric off-diagonal interaction, which is different from the $\Gamma$ term in (\ref{eq:H_KJG}), arises as 
${\cal H}_{\Gamma^\prime} = \Gamma' \sum_\gamma \sum_{\bar{\gamma}\neq \gamma} (S_i^\gamma S_j^{\bar{\gamma}} + S_i^{\bar{\gamma}} S_j^\gamma)$ 
with $\Gamma'>0$.

\section{Kitaev candidates beyond the Jackeli-Khaliullin mechanism}
\label{sec:beyond_JK}

In sections~\ref{sec:j=1/2} and \ref{sec:exch}, we discussed the electron configurations which can host the effective $j_{\rm eff}=1/2$ moments and the dominant exchange interactions between them. 
In the $d$-electron case, an interesting case is the high-spin $d^7$ electron configuration, which may provide dominant FM Kitaev interactions as the low-spin $d^5$ case in the Jackeli-Khaliullin mechanism. 
On the other hand, in the $f$-electron case, while there are several interesting electron configurations, we focused on the $4f^1$ case that can provide dominant AFM Kitaev interactions. 
In this section, we showcase the candidate materials for these two cases (sections~\ref{sec:d7} and \ref{sec:f1}). 
We also discuss another pathway to the AFM Kitaev interaction by introducing polar asymmetry in the lattice structure (section~\ref{sec:polar}).

\subsection{High-spin $d^7$ systems}
\label{sec:d7}

Among three magnetic ions with the high-spin $d^7$ electron configuration listed in table~\ref{table_d}, the prime candidate is Co$^{2+}$, as there are several compounds which have quasi-two-dimensional honeycomb structures composed of the edge-sharing network of CoO$_6$ octahedra. 
For instance, Na$_2$Co$_2$TeO$_6$ and Na$_3$Co$_2$SbO$_6$ were synthesized in a double-layer hexagonal structure with space group $P6_322$ and a single-layer monoclinic structure with space group $C2/m$, respectively~\cite{Viciu2007}. 
Their magnetic properties were studied in detail for powder samples~\cite{Lefrancois2016,Wong2016,Bera2017} and single crystals~\cite{Yan2019,Xiao2019}. 
Both compounds exhibit a zigzag-type AFM order at low temperature, similar to the low-spin $d^5$ candidates, Na$_2$IrO$_3$~\cite{Singh2010} and $\alpha$-RuCl$_3$~\cite{Johnson2015}. 
While the stability of the zigzag order was discussed by the Heisenberg model with further-neighbor couplings~\cite{Lefrancois2016}, the importance of the Kitaev interaction was pointed out based on the mechanism in sections~\ref{sec:j=1/2_d} and \ref{sec:exch_d}~\cite{Liu2018}. 
The Curie-Weiss behavior at high temperature indicates larger values of the effective magnetic moments than that expected from spin only, suggesting orbital contributions through the spin-orbit coupling, but they are also larger than the value for the effective $j_{\rm eff}=1/2$ moment~\cite{Yan2019}. 
Furthermore, a density-functional-theory calculation shows strong covalency between Co $d$ and O $p$ orbitals, which may oppose the formation of the $j_{\rm eff}=1/2$ doublet~\cite{Yan2019}. 
Further studies are needed for identifying the importance of the Kitaev interactions in these materials. 
We note that related materials were also studied recently, such as Ag$_3$Co$_2$SbO$_6$~\cite{Zvereva2016} and Li$_3$Co$_2$SbO$_6$~\cite{Stratan2019,Brown2019}. 

Other honeycomb candidates with Co$^{2+}$ cations are BaCo$_2$(AsO$_4$)$_2$~\cite{Regnault1977,Regnault1979,Regnault2018} and BaCo$_2$(PO$_4$)$_2$~\cite{Nair2018}. 
Both compounds crystalize in a rhombohedral structure with space group $R\bar{3}$ composed of a stack of undistorted honeycomb layers. 
At low temperature, BaCo$_2$(AsO$_4$)$_2$ shows a quasi-collinear order with a staggered out-of-plane component~\cite{Regnault2018}, while BaCo$_2$(PO$_4$)$_2$ exhibits a helical order~\cite{Nair2018}. 
For the latter compound, a partial substitution of P by V leads to spin-glass behavior, for which a transition to a nonmagnetic state was observed by applying a magnetic field~\cite{Zhong2018}. 
Meanwhile, for the former compound, it was shown recently that a magnetic field of $0.5$~T can suppress the magnetic order and induce a nonmagnetic state~\cite{Zhong2019preprint}. 
This is similar to the field-induced transition found in $\alpha$-RuCl$_3$~\cite{Kubota2015,Johnson2015}, where a possible Kitaev QSL has been intensively discussed in the field-induced state~\cite{Banerjee2018,Kasahara2018,Yokoi2020preprint}. 
It is worth noting that the critical fields are considerably small compared to $\sim 8$~T for $\alpha$-RuCl$_3$. 
Thus, these findings will stimulate further experiments to identify the nature of the field-induced states in these Co$^{2+}$-based compounds. 

Further potential candidates are an ilumenite CoTiO$_3$~\cite{Ishikawa1958,Newnham1964,Osmond1964,Lin2006,Yuan2019preprint} and a transition metal trichalcogenide CoPS$_3$~\cite{Wildes2017}. 
For the former, an interesting Dirac-like magnon dispersion was observed in the $A$-type AFM ordered phase (intraplane FM and interplane AFM)~\cite{Yuan2019preprint}. 
Meanwhile, the latter compound shows a zigzag-type AFM order, similar to other materials mentioned above. 
Also, Co$_4$Nb$_2$O$_9$ and Co$_4$Ta$_2$O$_9$ have honeycomb layers of Co$^{2+}$ bridged by corner-shared CoO$_6$ octahedra~\cite{Bertaut1961}, in which, however, the magnetoelectric properties have been recently attracted attention~\cite{Fang2015,Khanh2016,Solovyev2016,Khanh2017,Yanagi2018a,Yanagi2018b,Chaudhary2019}. 

It will also be intriguing to explore the Kitaev-type interactions in the Co$^{2+}$-based compounds on other lattice structures~\cite{Liu2018}. 
The candidates include quasi-two-dimensional triangular lattice compounds Ba$_3$CoSb$_2$O$_9$~\cite{Zhou2012}, Ba$_8$CoNb$_6$O$_{24}$~\cite{Rawl2017}, and Ba$_2$La$_2$CoTe$_2$O$_{12}$~\cite{Kojima2018}, and three-dimensional compounds, such as a spinel GeCo$_2$O$_4$~\cite{Tomiyasu2011,Pramanik2019} and pyrochlores NaCaCo$_2$F$_7$ and NaSrCo$_2$F$_7$~\cite{Krizan2015,Ross2016,Ross2017,Frandsen2017}. 

Besides Co$^{2+}$, Ni$^{3+}$ is also a candidate for the high-spin $d^7$ magnetic ion as listed in table~\ref{table_d}. 
Indeed, a layered honeycomb compound NaNi$_2$BiO$_{6-\delta}$~\cite{Seibel2014} was recently discussed in this context~\cite{Scheie2019}; 
the peculiar counterrotating magnetic order in the in-plane component was speculated to originate from Kitaev-type bond-dependent interactions between the spin-orbital entangled moments in the high-spin Ni$^{3+}$ state. 

Table~\ref{table_d7} summarizes the honeycomb materials with Co$^{2+}$ or Ni$^{3+}$ ions. 
The structural and magnetic properties are also shown. 

\begin{table}
\caption{Materials with honeycomb layers of edge-sharing ligand octahedra capsuling Co$^{2+}$ or Ni$^{3+}$ ions. Their structural and magnetic properties are shown: $T_{\rm m}$, $\mu_{\rm eff}$, and $\Theta_{\rm CW}$ are the magnetic transition temperature, the effective magnetic moment, and the Curie-Weiss temperature, respectively. The multiple numbers for $T_{\rm m}$ indicate successive transitions. AF denotes antiferromagnetic; ($c$) and ($ab$) indicate the values in a magnetic field applied along the $c$ axis and in the $ab$ plane, respectively.}
\label{table_d7}
\centering
\begin{tabular}{ccccccl}
\br
material & \begin{tabular}{c} space \\ group \end{tabular} & $T_{\rm m}$ (K) & \begin{tabular}{c} magnetic \\ ground state \end{tabular} & $\mu_{\rm eff}$ ($\mu_{\rm B}$) & $\Theta_{\rm CW}$ (K) & references \\ 
\mr
Na$_2$Co$_2$TeO$_6$ & $P6_322$ & $27$, $16$, $4$ & zigzag & $5.98$ ($c$) & $-139$ ($c$) & \cite{Lefrancois2016,Viciu2007,Bera2017,Xiao2019} \\
 & & & & $5.99$ ($ab$) & $-9$ ($ab$) & \\
Na$_3$Co$_2$SbO$_6$ & $C2/m$ & $5$ & zigzag & $5.60$ ($c$) & $-170$ ($c$) & \cite{Viciu2007,Wong2016,Yan2019} \\
 & & & & $5.48$ ($ab$) & $-4$ ($ab$) & \\
Ag$_3$Co$_2$SbO$_6$ & $C2/m$ & $21.2$ & - & $6.7$  & $-9$ & \cite{Zvereva2016} \\
Li$_3$Co$_2$SbO$_6$ & $C2/m$ & $14$ & A-type AF & $5.04$ & $18.1$ & \cite{Stratan2019,Brown2019} \\
BaCo$_2$(AsO$_4$)$_2$ & $R\bar{3}$ & $5.4$ & canted AF & $6.9$ ($c$) & $35$ ($c$) & \cite{Regnault1977,Regnault1979,Regnault2018,Zhong2019preprint} \\
 & & & & $7.4$ ($ab$) & $-90$ ($ab$) & \\
BaCo$_2$(PO$_4$)$_2$ & $R\bar{3}$ & $6$, $3.5$ & noncollinear & - & -& \cite{Nair2018} \\
CoTiO$_3$ & $R\bar{3}$ & $38$ & A-type AF & $5.3$ & $-15$ & \cite{Ishikawa1958,Newnham1964,Osmond1964,Lin2006,Yuan2019preprint} \\
CoPS$_3$ & $C2/m$ & $122$ & zigzag & $4.55$ & $-9.2$ ($c$) & \cite{Wildes2017} \\
 & & & & & $-87.9$ ($ab$) & \\
Co$_4$Nb$_2$O$_9$ & $P\bar{3}c1$ & $27.4$ & canted AF & $5.2$ ($c$) & $132$ & \cite{Bertaut1961,Khanh2016,Solovyev2016,Chaudhary2019} \\
 & & & & $5.1$ ($ab$) & $24$ ($ab$) & \\
Co$_4$Ta$_2$O$_9$ & $P\bar{3}c1$ & $20.5$ & - & $5.3$ & $-60$ & \cite{Bertaut1961,Solovyev2016,Fang2015} \\
NaNi$_2$BiO$_{6-\delta}$ & $P\bar{3}1m$ & $6.3$, $4.8$ & noncollinear & $2.21$ & $-18.5$ &\cite{Seibel2014,Scheie2019} \\
\br 
\end{tabular}
\end{table}

\subsection{$4f^1$ systems}
\label{sec:f1}

In the case of the $4f^1$ electron configuration, possible magnetic ions are Ce$^{3+}$ and Pr$^{4+}$, as listed in table~\ref{table_f}. 
Although no Ce$^{3+}$-based candidates are known to the best of our knowledge, several polymorphs are available in Pr$^{4+}$-based materials, $A_2$PrO$_3$ ($A$= alkali metals). 
For instance, Li$_2$PrO$_3$ crystalizes in an orthorhombic structure with space group of $Cmmm$~\cite{Wolf1987,Hinatsu2006}. 
This structure consists of quasi-one-dimensional chains of edge-sharing PrO$_6$ octahedra. 
When the mechanism discussed in sections~\ref{sec:j=1/2} and \ref{sec:exch} applies, strong Ising-like anisotropy is expected from the dominant AFM Kitaev interaction. 
Although such anisotropy was not studied since a single crystal is not available thus far, the magnetic susceptibility for powder samples indicates the existence of the effective $j_{\rm eff}=1/2$ moment and a magnetic order at low temperature~\cite{Hinatsu2006} (the ordered structure has not been identified yet). 

On the other hand, Na$_2$PrO$_3$ has a monoclinic structure with space group of $C2/c$~\cite{Hinatsu2006}. 
This is a quasi-two-dimensional structure with honeycomb layers of edge-sharing PrO$_6$ octahedra, similar to the iridium oxides Na$_2$IrO$_3$ and $\alpha$-Li$_2$IrO$_3$~\cite{Singh2010,Singh2012} 
(a partial mixing of Na and Pr was observed~\cite{Hinatsu2006}). 
In this case also, the susceptibility measurement indicates the formation of the effective $j_{\rm eff}=1/2$ moment and a magnetic order at low temperature with bifurcation of the susceptibility between the zero-field-cooled and field-cooled data~\cite{Hinatsu2006}. 

A different quasi-two-dimensional structure with a stack of triangular layers (space group $R\bar{3}m$) was found in Na$_2$PrO$_3$~\cite{Brunn1977} and K$_2$PrO$_3$~\cite{Paletta1966,Brunn1977}. 
In addition, a three-dimensional structure with space group $C2/c$ was synthesized for Na$_2$PrO$_3$~\cite{Wolf1988}, which has the hyperhoneycomb-type network of edge-sharing CoO$_6$ octahedra similar to $\beta$-Li$_2$IrO$_3$~\cite{Takayama2015}. 
For other $A$-site alkali ions, Rb and Cs, there were no experimental reports to the best of our knowledge, but a hyperhoneycomb-type structure with space group $Fddd$ was proposed by {\it ab initio} calculations for Rb$_2$PrO$_3$~\cite{Persson2016}. 

The experimentally-synthesized polymorphs are summarized in table~\ref{table_A2PrO3}. 
The known structural and magnetic properties are also shown. 

\begin{table}
\caption{Synthesized polymorphs of $A_2$PrO$_3$ with edge-sharing networks of ligand octahedra capsuling Pr$^{4+}$. Their structural and magnetic properties are also shown by the common notations to table~\ref{table_d7}.}
\label{table_A2PrO3}
\centering
\begin{tabular}{cccccccl}
\br
material & \begin{tabular}{c} space \\ group \end{tabular} & Pr network & $T_{\rm m}$ (K) & $\mu_{\rm eff}$ ($\mu_{\rm B}$) & $\Theta_{\rm CW}$ (K) & references \\ 
\mr
Li$_2$PrO$_3$ & $Cmmm$ & chain & $6.5$ & $1.75$ & $-32$ & \cite{Wolf1987,Hinatsu2006} \\
Na$_2$PrO$_3$ & $C2/c$ & honeycomb & $4.6$ & $0.99$ & $-15$ & \cite{Hinatsu2006} \\
 & $R\bar{3}m$ & triangular & - & - & - & \cite{Brunn1977} \\
 & $C2/c$ & hyperhoneycomb & - & - & - & \cite{Wolf1988} \\
K$_2$PrO$_3$ & $R\bar{3}m$ & triangular & - & $2.40$ & $133$ & \cite{Paletta1966,Brunn1977} \\
\br 
\end{tabular}
\end{table}

As mentioned in the end of section~\ref{sec:exch_f}, a series of $A_2$PrO$_3$ was theoretically studied by combining {\it ab initio} calculations and model analyses~\cite{Jang2019,Jang2019preprint}. 
In particular, the quasi-two-dimensional honeycomb forms of the compounds were studied systematically for the $A$ cations, Li, Na, K, Rb, and Cs. 
The structural optimization by the {\it ab initio} calculations show that all the compounds converge onto monoclinic structures with $C2/m$ symmetry, and the electronic band structure indicates the formation of the effective $j_{\rm eff}=1/2$ moments in the $\Gamma_7$ doublet, as expected from the results in section~\ref{sec:j=1/2_f}. 
It was shown that, for larger $A$-site ionic radii, trigonal distortions of the PrO$_6$ octahedra as well as the distances between neighboring Pr cations are enhanced, which introduce larger deviations from the ideal situation with the dominant AFM Kitaev interactions discussed in section~\ref{sec:exch_f}. 
This indicates that Li$_2$PrO$_3$ will be the best candidate for the Kitaev magnet in this series, although the honeycomb-type polymorph has not been synthesized yet~\cite{Wolf1987,Hinatsu2006}. 
Na$_2$PrO$_3$ is also a good candidate, as it was already synthesized experimentally. 
For Na$_2$PrO$_3$, the theoretical study was extended to the three-dimensional hyperhoneycomb form~\cite{Jang2019preprint}, which was also synthesized~\cite{Wolf1988}, and the dominant AFM Kitaev interaction is expected also in this case.

\subsection{Polar asymmetric systems}
\label{sec:polar}

In the $4f^1$ case, the dominant AFM Kitaev interaction is generated by the spatial anisotropy of the $f$ orbitals and the small crystal field splitting, as discussed in section~\ref{sec:exch_f}. 
Another way to induce the AFM Kitaev interaction, which can work for the $d$-electron cases, was proposed theoretically by some of the authors~\cite{Sugita2019preprint}. 
This was discussed for the honeycomb structure composed of the low-spin $d^5$ cations, but potentially applicable to other tricoordinate structures and the high-spin $d^7$ cations since the mechanism for realizing the Kitaev interactions is essentially the same as discussed in sections~\ref{sec:j=1/2_d} and \ref{sec:exch_d}. 
Let us outline the mechanism below. 

Suppose the honeycomb structure is modulated in an asymmetric way perpendicular to the plane. 
Such a situation may be realized, for instance, on a surface of quasi-two-dimensional layered structures, a heterostructure with other materials, and partial substitution of the ligand ions. 
The polar asymmetry hampers the quantum interference between two $d$-$p$-$d$ paths in the requisite (ii) in the Jackeli-Khaliullin mechanism. 
The most relevant contribution arises from a ``Rashba-type" hopping, which is spin and orbital dependent and has an imaginary matrix element. 
The exchange processes through this hopping between the $j_{\rm eff}=1/2$ states lead to the Kitaev and Heisenberg coupling constants~\cite{Sugita2019preprint} 
\begin{equation}
K \sim - \frac{8}{U} \tilde{\eta}^2, 
\quad
J \sim - \frac{4}{U} \tilde{\eta}^2, 
\end{equation}
where $\tilde{\eta}$ is the amplitude of the Rashba-type spin-dependent hopping. 
The important point in this mechanism is that 
(i) the Kitaev coupling originates from the perturbation process between the $j_{\rm eff}=1/2$ states, while that in the Jackeli-Khaliullin mechanism is from the perturbation via the higher-energy $j_{\rm eff}=3/2$ state, and 
(ii) $K$ is negative (AFM) and proportional to $1/U$, in contrast to the FM one proportional to $J_{\rm H}/U^2$ in the Jackeli-Khaliullin mechanism [see (\ref{eq:K_d5})]. 
The latter is particularly important since it suggests the possibility of larger Kitaev couplings compared to the conventional mechanism. 

The idea was tested by combining {\it ab initio} calculations and model analyses~\cite{Sugita2019preprint}. 
As a representative situation, starting from a Kitaev candidate $\alpha$-RuCl$_3$, monolayers of  $\alpha$-RuH$_{3/2}X_{3/2}$ ($X$=Cl and Br) are considered, where the polar asymmetry is introduced by replacing the halides on one side of the honeycomb layer by hydrogens. 
Note that similar polar structures were indeed fabricated in related transition metal compounds~\cite{Lu2017,Kageyama2018}. 
As expected from the above arguments, the AFM Kitaev couplings $K$ were obtained, whose amplitudes are several times larger than the FM one for $\alpha$-RuCl$_3$. 

As mentioned above, similar polar asymmetry will be seen in more generic cases. 
An interesting situation is a surface or an interface of the Kitaev candidate materials. 
For instance, a van der Waals material $\alpha$-RuCl$_3$ was successfully fabricated in a thin film form~\cite{Weber2016,Ziatdinov2016,Gronke2018,Mashhadi2018,Tian2019}. 
Their surfaces will provide a testbed for the above mechanism. 
Furthermore, recently, heterostructures between $\alpha$-RuCl$_3$ and graphene have attracted a lot of attention for peculiar electronic properties potentially including information on the exotic magnetism in $\alpha$-RuCl$_3$~\cite{Zhou2019,Mashhadi2019,Biswas2019,Gerber2020}. 
In these systems, polar asymmetry is inherently present, and hence, the above mechanism may induce large AFM Kitaev interactions in $\alpha$-RuCl$_3$ near the interface.

\section{Summary and perspective}
\label{sec:summary}

To summarize, we have overviewed the recent theoretical exploration of the Kitaev magnets beyond the Jackeli-Khaliullin mechanism. 
The two requisites, formation of the $j_{\rm eff}=1/2$ doublet and quantum interference between different indirect hoppings, were carefully reexamined for both $d$- and $f$-electron cases. 
First, we presented the systematic analysis of the electron configurations which can host the $j_{\rm eff}=1/2$ Kramers doublet. 
In the $d$-electron case, the high-spin $d^7$ state is nominated as the candidate, in addition to the low-spin $d^5$ one known in the Jackeli-Khaliullin mechanism. 
Meanwhile, in the $f$-electron case, there are several candidates: the $\Gamma_7$ doublet for the $4f^1$ and $4f^5$ states, and the $\Gamma_6$ doublet for the $4f^9$ and $4f^{13}$ states. 
The $4f^3$ $\Gamma_6$ and $4f^{11}$ $\Gamma_7$ states can be the candidates as well, depending on the crystal field splitting. 
We also pointed out that the $5f^1$ case has a similar $\Gamma_7$ state to $4f^1$. 
Next, we discussed the exchange processes between these $j_{\rm eff}=1/2$ moments for the edge-sharing octahedra. 
For the $d$-electron case, we discussed the exchange processes for the high-spin $d^7$ state, which lead to the dominant FM Kitaev interaction as in the low-spin $d^5$ case in the Jackeli-Khaliullin mechanism. 
On the other hand, among the several candidates in the $f$-electron case, we focused on the $4f^1$ electron configuration, where the dominant AFM Kitaev interaction arises from the peculiar spatial anisotropy of the $f$ orbitals and the small crystal field splitting. 
Based on these observations, we discussed the candidate materials beyond the Jackeli-Khaliullin mechanism. 
For the high-spin $d^7$ case, we listed several Co$^{2+}$- and Ni$^{3+}$-based materials. Meanwhile, for the $4f^1$ case, we nominated polymorphs of Pr$^{4+}$ based materials. 
In addition, we discussed another candidate with structural polar asymmetry, which leads to the dominant AFM Kitaev interaction even in the low-spin $d^5$ and high-spin $d^7$ cases. 

These progresses will stimulate further material exploration of the Kitaev spin liquids. 
There remain many unexplored issues in both $d$- and $f$-electron systems. 
For instance, it will be intriguing to examine whether the Kitaev interaction is relevant to the magnetic properties in the Co$^{2+}$- and Ni$^{3+}$-based compounds discussed in section~\ref{sec:d7}. 
In particular, the field-induced nonmagnetic states found in BaCo$_2$(AsO$_4$)$_2$ and BaCo$_2$(P$_{1-x}$V$_x$)O$_8$ will attract much attention in comparison with that in the $d^5$ candidate $\alpha$-RuCl$_3$. 
In the $f$-electron case, it is worth trying to synthesize the polymorphs of $A_2$PrO$_3$ ($A$: alkali metals) in the quasi-two- and three-dimensional forms, and measure their magnetic properties to find the signatures of the Kitaev spin liquids. 
In addition, compounds with $4f^5$ (Sm$^{3+}$), $4f^9$ (Dy$^{3+}$), $4f^{13}$ (Tm$^{2+}$ and Yb$^{3+}$), and $5f^1$ (Pa$^{4+}$ and U$^{5+}$) would be worth investigating. 
We note that some efforts have been done toward the Kitaev physics in the $f$-electron systems, such as double perovskites~\cite{Li2017,Luo2019preprint} and pyrochlores~\cite{Thompson2017,Rau2019}. 
The polar asymmetry will also be a relevant issue to the recent development in thin films of the Kitaev magnets and also to future development in electronic and magnetic devices. 
Finally, we emphasize again that the AFM Kitaev interactions, which are expected for the $4f^1$-electron systems and the polar asymmetric $d$-electron systems, are crucially important as they are not realized in the existing low-spin $d^5$ candidates. 
They will enable us to access the unexplored parameter regions where a different QSL state from the Kitaev one is theoretically anticipated in a magnetic field~\cite{Zhu2018,Gohlke2018,Nasu2018,Liang2018,Hickey2019,Ronquillo2019}.

\ack
The authors thank 
R. Coldea, G. Khaliullin, K. Matsuhira, T. Miyake, J. Nasu, K. Nomura, H. Shinaoka, A. Tsukazaki, Y. Yamaji, and J. Yoshitake for fruitful discussions. 
Y.S. was supported by the Japan Society for the Promotion of Science through a research fellowship for young scientists and the Program for Leading Graduate Schools (MERIT). 
This work was supported by 
JSPS KAKENHI Grant Nos. JP24340076, JP15K13533, JP16H02206, and JP18K03447, JST CREST (JP-MJCR18T2), and US NSF PHY-1748958. 

\vspace{8mm}

\bibliographystyle{iopart-num}
\bibliography{ref}

\end{document}